\documentclass[doublecol]{epl2}
\usepackage{amsmath}

\title{Continuous variable quantum key distribution based on optical
entangled states without signal modulation}\shorttitle{CV QKD based
on optical entangled states}
\author{Xiaolong Su, Wenzhe Wang, Yu Wang, Xiaojun Jia, Changde Xie\thanks{%
E-mail: changde@sxu.edu.cn}, and Kunchi Peng}\shortauthor{Xiaolong
Su\etal}
\institute{The State Key Laboratory of Quantum Optics and Quantum Optics Devices,\\
Institute of Opto-Electronics, Shanxi University, Taiyuan, 030006,
P.R.China } 
 \abstract{In this paper, we present the first
experimental demonstration on continuous variable quantum key
distribution using determinant Einstein-Podolsky-Rosen entangled
states of optical field. By means of the instantaneous measurements
of the quantum fluctuations of optical modes respectively
distributed at sender and receiver, the random bits of secret key
are obtained without the need for signal modulation. The
post-selection boundaries for the presented entanglement-based
scheme against both Gaussian collective and individual attacks are
theoretically concluded. The final secret key rates of 84 kbits/s
and 3 kbits/s are completed under the collective attack for the
transmission efficiency of 80\% and 40\%, respectively.}

\begin{document}

\maketitle

Quantum key distribution (QKD) allows two legitimate parties, Alice
and Bob, to establish the secret key only known by themselves. A
secret key is usually generated by Alice transmitting the prepared
quantum states to Bob, who performs measurements on the received
states to distill the information. There are two types of QKD
systems in which the discrete or continuous quantum variables are
exploited, respectively. For discrete variable (DV) QKD protocols
the key information is encoded in discrete quantum variables of
single photon light pulse, such as polarization or phase
\cite{Gisin}. In continuous variable (CV) QKD protocols continuous
quantum variables of light field, such as amplitude and phase
quadratures, are used for transmitting information. Comparing with
DV QKD of single photon schemes CV QKD promises significantly higher
secret key rates and eliminates the need for single photon
technology. Recently, coherent state CV QKD protocols have been
experimentally demonstrated \cite{Gros1,Lorenz,Lance,Lod,Qi}. These
successful experiments proved that CV QKD is a hopeful and viable
path to develop quantum cryptography for real-world applications. On
the other hand the strictly theoretical proofs on the security of CV
QKD protocols using
both coherent and non-classical states of light have been achieved \cite%
{Got,Ibl,Gros2,Nav1}. CV QKD protocols have recently been shown to
be unconditionally secure, that is, secure against arbitrary attacks
\cite{Ren} and have been proved to be unconditionally secure over
long distance \cite{Lev}.

Quantum entanglement is one of the quite essential features in
quantum mechanics that has no analogue in classical physics. It has
been theoretically demonstrated by Curty \textit{et al.} that the
presence of detectable entanglement in a quantum state effectively
distributed between sender (Alice) and receiver (Bob) is a necessary
precondition for successful key distillation \cite{Curty}. However,
there is no CV QKD experiment directly utilizing optical entangled
states to be presented until now, although a variety of theoretical
CV QKD protocols based on Einstein-Podolsky-Rosen (EPR) entanglement
and squeezing of optical fields have been proposed
\cite{Ralph,Hil,Cerf,Reid,Ben,Sil,Su,Pir}. Not like CV
QKD protocols applying coherent states of light \cite%
{Gros1,Lorenz,Lance,Lod,Qi}, in which the bits of secret key are
constructed classically using amplitude and phase modulation,
so-called prepare-and-measure (P\&M) scheme \cite{Lod}, in the
entanglement-based (EB) schemes proposed by refs. 16 and 17 the bits
of the random secret key are constructed by the instantaneous
measurements of the correlated quantum fluctuations of the
quadratures between two entangled optical modes distributed at Alice
and Bob. In the EB CV QKD protocols, the quantum fluctuations of
entangled optical beams with the truly quantum randomness are
utilized to generate the key. Due to that the classical signal
modulation is not needed, the bit rates will not be limited by the
rates of the electronic modulators and the experimental systems will
be simplified.

In the presented paper, we experimentally demonstrated the
proof-of-principle CV QKD protocol using a pair of bright EPR
entangled beams produced from a non-degenerate optical parametric
amplifier (NOPA). We concluded the post-selection boundaries of the
presented EB CV QKD scheme against both Gaussian collective and
individual attacks. By means of the post-selection, reconciliation
and privacy amplification techniques, the final secret key was
obtained through distilling the measured data of the correlated
quantum fluctuations of quadratures. The generated raw key rate is 2
Mbits/s and the final secret key rates are 84 kbits/s and 3 kbits/s
against Gaussian collective attack for the transmission efficiency
of 80\% and 40\%, respectively. We believe that this is the first
experimental demonstration of CV QKD protocols directly exploiting
the EPR entanglement of amplitude and phase quadratures of optical
field. On the physical sense this experiment intuitionally shows the
close relationship between the security of CV QKD and the quantum
entanglement.

The experimental setup of the CV QKD protocol is shown in fig. 1.
The laser is a homemade continuous wave intracavity
frequency-doubled and frequency stabilized Nd:YAP/KTP ring laser
consisting of five mirrors \cite{Jia}. The second harmonic wave
output at 540 nm is used for the pump field of the NOPA and the
fundamental wave output at 1080 nm is separated into two parts, one
is for the injected signal of the NOPA and the other is used as the
local oscillation beams of the homodyne detections for Alice and
Bob. The NOPA consists of an $\alpha $-cut type-II KTP crystal and a
concave mirror. Through a parametric down conversion process of type
II phase match, a pair of EPR beams with anticorrelated amplitude
quadratures and correlated phase quadratures may be produced from
the NOPA operating in the state of de-amplification, that is, the
pump field and the injected signal are out of phase \cite{Li}. The
bandwidth of the NOPA is about 20 MHz, in which the output beams are
entangled. If distributing the two beams of EPR pair to Alice (beam
a) and Bob (beam b), the instantaneous measurement outcomes of
quadrature quantum fluctuations on their respective modes will be
fairly identical due to the quantum correlations of quadratures
\cite{Takei}.

In the communication, Alice and Bob randomly measure the amplitude
or phase quadrature of the entangled optical beam they hold
respectively, with the homodyne detection systems. After the
measurement is completed, they compare the measurement basis in the
authorized classic channel and only remain the measurement results
of the compatible basis. Then they use post-selection technique to
select a subset from the measured raw data to make the mutual
information of Alice and Bob advantage over Eve's information. To
implement the post-selection, Alice publicly announce the absolute
values of the measured amplitude or phase quadratures ($|X_{A}|$ or
$|Y_{A}|$), but not publicly open their symbols \cite{Lance}. Alice
and Bob also choose a random subset of data to characterize the
channel efficiency and excess noise. From these values they select
the secure data and discard the insecure data. After post-selection
procedure, Alice and Bob interpret the post-selected data into
binary data. For the correlated phase quadratures both Alice and Bob
may define the positive and negative phase fluctuations as a binary
\textquotedblleft 1\textquotedblright\ and \textquotedblleft
0\textquotedblright , respectively. However for the anticorrelated
amplitude quadratures if Alice define the positive (negative)
amplitude fluctuation as \textquotedblleft 1\textquotedblright\
(\textquotedblleft 0\textquotedblright ), Bob should defines
negative (positive) amplitude fluctuation as \textquotedblleft
1\textquotedblright\ (\textquotedblleft 0\textquotedblright ). Then
we apply the reconciliation protocol to correct the errors of the
retained data. At last, we apply a privacy amplification procedure
to distill the final secret key.

\begin{figure}[tbp]
\begin{center}
\includegraphics[width=8cm,height=5cm]{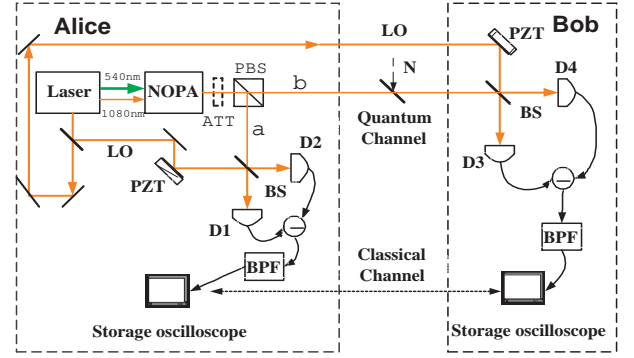}
\end{center}
\caption{The experimental system of QKD. NOPA: non-degenerate
optical parametric amplifier, ATT: attenuator, PBS: polarization
beam splitter, LO: local oscillation beam, N: vacuum noise, PZT:
piezoelectric transducer, BS: 50/50 beam splitter, D1-D4:
photodetector, BPF: band-pass filter.}
\end{figure}

In the security analysis we assume that the quantum channel
connecting Alice and Bob is lossy with imperfect transmittivity of
$\eta $ and the Gaussian excess noise $\delta $\ on the quadrature
distribution exists in the communication system. The security
analyses are restricted to protect
against Gaussian attacks only. For the optimal beam splitter attack \cite%
{Gros4}, that is, Eve takes a fraction $1-\eta $\ of the beam b at
Alice's site and sends the fraction $\eta $\ to Bob through her own
lossless line. In this case Eve is totally undetected, and she gets
the maximum possible information according to the no-cloning
theorem. For collective attack,\ Eve listens to the communication
between Alice and Bob during the key distillation procedure and then
applies the optimal collective measurement on the ensemble of stored
ancilla. The maximum information Eve may have access to is limited
by the Holevo bound $\chi $ \cite{Lod}. Under the individual\
attack, Eve measures the intercepted ensemble before the key
distillation stage and Eve's information is summarized by the mutual
information between Alice and Eve, $I_{AE}$, for direct
reconciliation. Generally, the information exchange is secure as
long as the mutual information between Alice and Bob ($I_{AB}$) is
larger than Eve's information. The condition of $I_{AB}>I_{AE}$ for
extracting secure key directly results in the restriction of the
maximum transmission losses less than 3 dB which will limit the
possible transmission distances \cite{Gros4}. Fortunately, the 3 dB
loss limit for CV QKD protocols can be beaten by implementing a
reverse reconciliation scheme \cite{Gros1} or applying an
appropriate post-selection \cite{Lorenz,Lance,Silber}. It has been
shown that there must be a lower limit of $\eta $ (2$\eta >\delta $)
for the secure key distillation if the excess noise exists
\cite{Namiki}. The security of EB scheme against collective attack
with reverse reconciliation has been proved \cite{Gros3}. Here, we
analyze the post-selection boundary for the EB scheme against
Gaussian collective attack and Gaussian individual attack.

For the EPR beams with anticorrelated amplitude quadratures and
correlated phase quadratures, we have the following relations
\cite{Gros3}:
\begin{eqnarray}
&\langle X_{a(b)}^{2}\rangle =\langle Y_{a(b)}^{2}\rangle
=VN_{0}=(e^{2r}+e^{-2r})N_{0}/2,& \\
&\langle (X_{a}+X_{b})^{2}\rangle =\langle (Y_{a}-Y_{b})^{2}\rangle
=2e^{-2r}N_{0},& \\
&\langle X_{a}X_{b}\rangle =-\sqrt{V^{2}-1}N_{0},& \\
\quad &\langle Y_{a}Y_{b}\rangle =\sqrt{V^{2}-1}N_{0},&
\end{eqnarray}%
where $r$\ is the correlation parameter, $N_{0}$\ = 1/4 is the
shot-noise-limited variance. These beam are entangled, and the
measurement of a quadrature of beam a (\textit{e.g.} $Y_{a}$) gives
Alice information on the same quadrature of the other beam
($Y_{b}$). By measuring the amplitude
quadrature $X_{a}$\ (phase quadrature $Y_{a}$) on her beam a, Alice learns $%
X_{A}$\ ($Y_{A}$), and projects the Bob's beam b onto a $X$-squeezed ($Y$%
-squeezed) state of squeezing parameter $s=1/V$\ centered on
($X_{A}$, 0) [(0, $Y_{A}$)] \cite{Gros3}. The best estimate Alice
can have on $Y_{b}$\
knowing $Y_{a}$ is of the form $Y_{A}=\alpha Y_{a}$\ with $\alpha =\frac{%
\langle Y_{b}Y_{a}\rangle }{\langle Y_{a}^{2}\rangle }$, the value of $%
\alpha $\ being found by minimizing the variance of the error operator $%
\delta Y_{A}=Y_{b}-Y_{A}$.\ The conditional variance $V_{Y_{b}\mid
Y_{A}}$\
of $Y_{b}$\ knowing $Y_{A}$\ quantifies the remaining uncertainty on $Y_{b}$%
\ after the measurement of $Y_{a}$\ giving the estimate $Y_{A}$\ of
$Y_{b}$, and we have

\begin{equation}
V_{Y_{b}\mid Y_{A}}=\langle \delta Y_{A}^{2}\rangle =\langle
Y_{b}^{2}\rangle -\frac{\left\vert \langle Y_{a}Y_{b}\rangle \right\vert ^{2}%
}{\langle Y_{a}^{2}\rangle }=\frac{N_{0}}{V}
\end{equation}%
Since by measuring $Y_{a}$\ Alice deduces $Y_{A}$, and since $%
Y_{b}=Y_{A}+\delta Y_{A}$, the beam $b$\ is projected onto a
$Y$-squeezed state with squeezing variance $V_{s}=V_{Y_{b}\mid
Y_{A}}=N_{0}/V$\ centered on $(0,Y_{A})$. Alternatively, by
measuring $X_{a}$, Alice learns $X_{A}$\
and projects the other beam onto a $X$-squeezed state centered on ($X_{A}$, $%
0$) with the same squeezing variance $V_{s}=N_{0}/V$. The variances
of quadratures measured by Alice and Bob are $V_{A}=\alpha
^{2}VN_{0}=(V-1/V)N_{0}$\ and $V_{B}=(\eta V_{A}+\eta V_{s}+1-\eta
+\delta )N_{0}$, respectively.

The probability that Bob obtains the measurement outcome $Y_{B}$ is given by%
\begin{equation}
P_{B}(Y|\Psi \rangle )=\frac{1}{\sqrt{2\pi V_{B}^{N}}}\exp [-\frac{(Y_{B}-%
\sqrt{\eta }Y_{A})^{2}}{2V_{B}^{N}}],
\end{equation}%
where $|\Psi \rangle $ represents the transmitted quantum state, the
noise variance $V_{B}^{N}=(\eta V_{s}+1-\eta +\delta )N_{0}$ of
which depends on the squeezed variance $\eta V_{s}N_{0}$, the
`vacuum noise' component due to the line losses $(1-\eta )N_{0}$,
and the `excess noise' component $\delta
N_{0}$. The corresponding Bob's error rate is given by%
\begin{equation}
p=\frac{P_{B}(Y|\Psi \rangle )}{P_{B}(Y|\Psi \rangle )+P_{B}(Y|-\Psi
\rangle )}=1/[1+\exp (\frac{4\sqrt{\eta
}Y_{A}|Y_{B}|}{2V_{B}^{N}})].
\end{equation}%
Based on eq. (7) we calculated the mutual information between Alice and Bob%
\begin{equation}
I_{AB}=1+p\log _{2}p+(1-p)\log _{2}(1-p).
\end{equation}

For collective attack, Eve's knowledge of the data can be quantified
by the Holevo bound $\chi $, which equals to \cite{Heid}

\begin{equation}
\chi =S(\overline{\rho })-\sum_{i=0}^{1}p_{i}S(\rho _{i}),\quad
\quad \rho =\sum_{i=0}^{1}p_{i}\rho _{i},
\end{equation}%
where $S(\rho )=-tr\rho \log _{2}\rho $ is the von Neumann entropy
of a quantum state $\rho $. The $\chi $ includes that Eve being
allowed to measure out her ancillas collectively. After Alice and
Bob have corrected their bit stings, Eve can use the information
transmitted over the public channel to optimize her measurements on
her ancilla systems. The quantum states in Eve's hand, conditioned
on Alice's data, are given by $|\Psi _{i}\rangle _{E}=|\pm
\sqrt{1-\eta }\Psi \rangle ,$ where $i=0,1$ denote the
encoded binary state. These states are pure, so that we have $\chi =S(%
\overline{\rho })$. What remains to be calculated are the eigenvalues of $%
\overline{\rho }=\frac{1}{2}(\left\vert \Psi _{0}\right\rangle
_{E}\left\langle \Psi _{0}\right\vert +\left\vert \Psi
_{1}\right\rangle _{E}\left\langle \Psi _{1}\right\vert )$. The
symmetry allows us to write
the states $\left\vert \Psi _{i}\right\rangle _{E}$ as%
\begin{align}
\left\vert \Psi _{0}\right\rangle _{E}& =c_{0}\left\vert \Phi
_{0}\right\rangle +c_{1}\left\vert \Phi _{1}\right\rangle \\
\left\vert \Psi _{1}\right\rangle _{E}& =c_{0}\left\vert \Phi
_{0}\right\rangle -c_{1}\left\vert \Phi _{1}\right\rangle  \notag
\end{align}%
where the $\left\vert \Phi _{i}\right\rangle $ are orthonormal
states. A short calculation shows that is already diagonal in this
basis with
eigenvalues $|c_{i}|^{2}$, so that the Holevo quantity is given by%
\begin{equation}
\chi =S(\overline{\rho })=-\sum_{i=0}^{1}|c_{i}|^{2}\log
_{2}|c_{i}|^{2}.
\end{equation}%
The normalization of $\rho $, $|c_{0}|^{2}+|c_{1}|^{2}=1$, and the overlap $%
|c_{0}|^{2}-|c_{1}|^{2}=_{E}\langle \Psi _{0}\left\vert \Psi
_{1}\right\rangle _{E}$ give the expressions for the coefficients,

\begin{align}
|c_{0}|^{2}& =\frac{1}{2}(1+_{E}\langle \Psi _{0}|\Psi _{1}\rangle _{E}), \\
|c_{1}|^{2}& =\frac{1}{2}(1-_{E}\langle \Psi _{0}|\Psi _{1}\rangle
_{E}). \notag
\end{align}%
The overlap of the two states can be calculated by $|_{E}\langle
\Psi
_{0}|\Psi _{1}\rangle _{E}|^{2}=\pi \int W(X,-Y)W(X,Y)dXdY$ if $Y$%
-quadrature is measured, where $W(X,Y)$ is the Wigner function of
the projected squeezed states centered on ($X_{0},Y_{0}$). If
Y-quadrature is measured by Alice, the correlation matrix of the
projected $Y$-squeezed
state can be written as%
\begin{equation}
\mathbf{V}_{c}=\frac{1}{4}\left[
\begin{array}{cc}
V & 0 \\
0 & 1/V%
\end{array}%
\right] .
\end{equation}

Using the expression of Wigner function for Gaussian states with one
dimensional vector \cite{Van}%
\begin{equation}
W(X,Y)=\frac{1}{2\pi \sqrt{\det \mathbf{V}_{c}}}\exp \{-\frac{1}{2}(X,Y)[%
\mathbf{V}_{c}]^{-1}(X,Y)^{T}\},
\end{equation}%
we can write out the corresponding Wigner function of the projected $Y$%
-squeezed state
\begin{equation}
W(X,Y)=\frac{2}{\pi }\exp [-\frac{4(X-X_{0})^{2}}{e^{2r}+e^{-2r}}%
-(e^{2r}+e^{-2r})(Y-Y_{0})^{2}].
\end{equation}%
If $Y$-quadrature are measured, then Eve's state is displaced to $Y_{0}=%
\sqrt{1-\eta }Y_{A}$, so the overlap between the two states $W(X,Y)$ and $%
W(X,-Y)$ is%
\begin{equation}
f=_{E}\langle \Psi _{0}|\Psi _{1}\rangle _{E}=\exp [-\frac{(1-\eta )Y_{A}^{2}%
}{2V_{s}}].
\end{equation}%
So, the Holevo quantity can be directly calculated. From eqs. (8)
and (11), we can obtain the secret key rates $K=I_{AB}-\chi $
against collective attack.

For the individual attack, the mutual information between Alice and
Eve is
expressed by \cite{Silber}%
\begin{align}
I_{AE}& =\frac{1}{2}(1+\sqrt{1-f^{2}})\log _{2}(1+\sqrt{1-f^{2}}) \\
& +\frac{1}{2}(1-\sqrt{1-f^{2}})\log _{2}(1-\sqrt{1-f^{2}}).  \notag
\end{align}%
The secret key rate against individual attack is $\Delta
I=I_{AB}-I_{AE}$. Of course, the security boundary can also be
directly applied to the anti-correlated amplitude quadrature X, for
that we only need to change the signs of the measured amplitude
values.

\begin{figure}[ptb]
\begin{center}
\includegraphics[width=7cm,height=6cm]{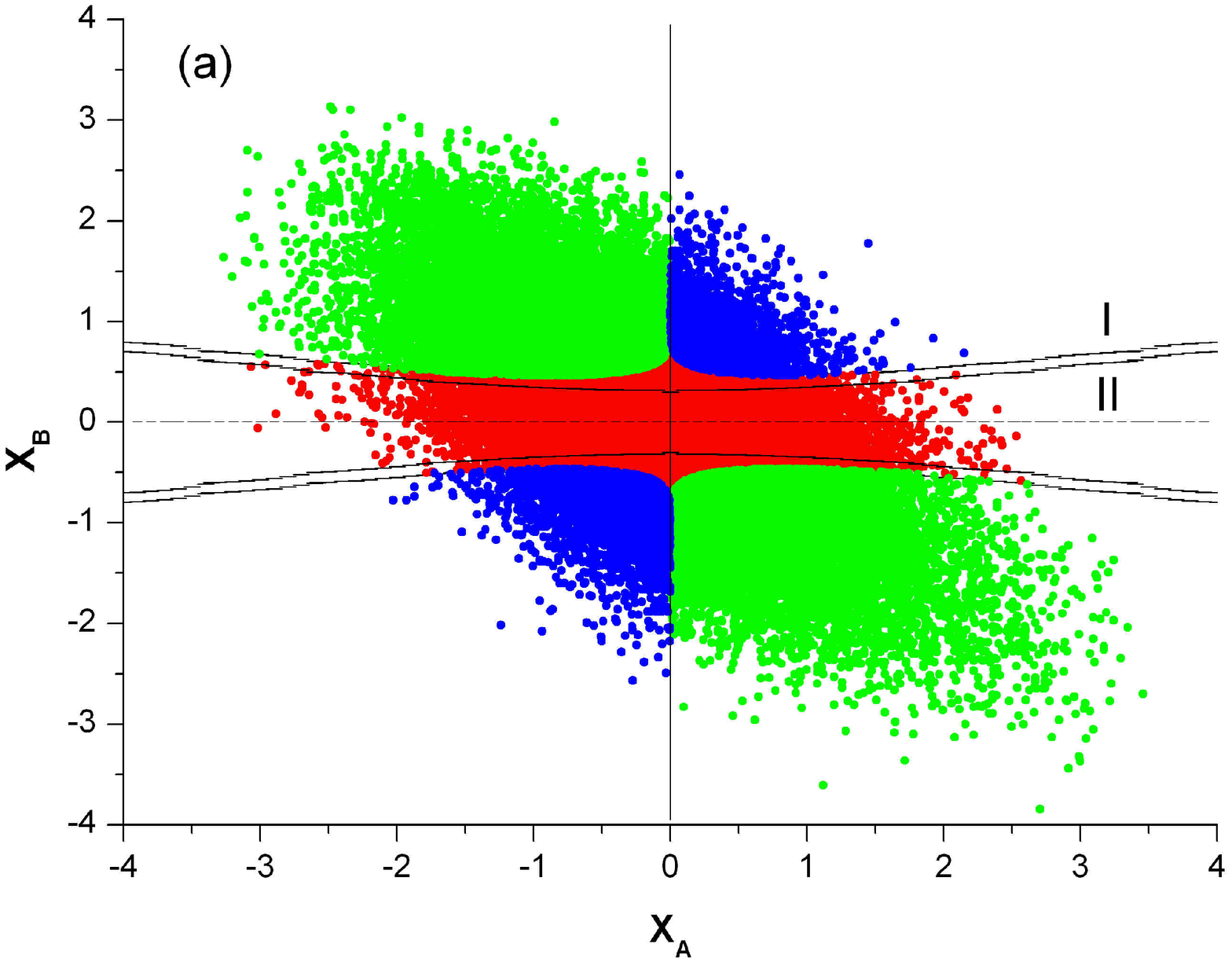} %
\includegraphics[width=7cm,height=6cm]{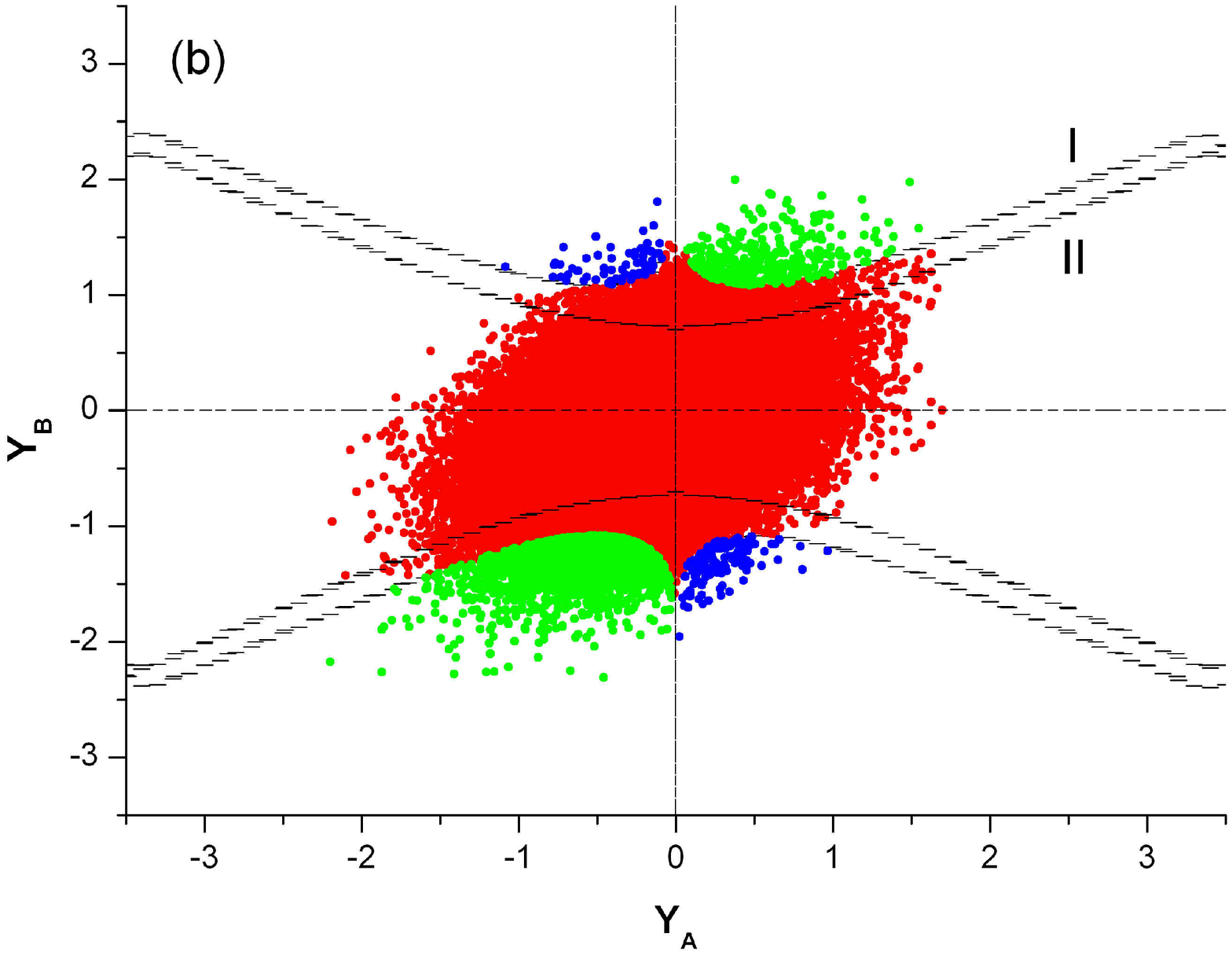}
\end{center}
\caption{The \textquotedblleft global\textquotedblright\ perspective
of Alice's and Bob's data. (a), Amplitude quadrature for 80\%
transmission efficiency. (b), Phase quadrature for 40\% transmmision
efficiency. I: post-selection boundary for collective attack. II:
post-selection boundary for individual attack. Green data points:
data that was error-free. Blue data points: data that has bit-flip
errors. Red data points: data that has a negative net information
rate. }
\end{figure}

In the communication, at first Alice separates the EPR entangled
beams generated by the NOPA with a polarized beam splitter (PBS) and
then sends one of them (beam b) to Bob while keeps the other one
(beam a) within her own station. The beam b is transmitted in air
about 2 meter. We simulated the QKD communication in two cases
respectively with the transmission efficiency of $80\%$ and $40\%$,
which were completed by inserting an appropriate attenuator into the
optical path. For making the balance attenuation of two optical and
implementing simultaneous measurements of the correlated quantum
fluctuations, we insert an attenuator (ATT) with transmission
efficiency 89\% or 45\% into the optical path of the EPR beams
before they are separated. In addition to the detection efficiency
of 90\%, the total transmission efficiency between Alice and Bob is
80\% ($89\%\times 90\%$) or 40\% ($45\%\times 90\%$), respectively.
During the communication, Alice and Bob randomly and instantaneously
measure the amplitude or phase quadratures of their own beam with a
homodyne detection system, which is completed by randomly switching
the phase difference between the local oscillation and the EPR beam
from 0 for the amplitude quadratures to $\pi /2$ for the phase
quadratures. The time interval $\Delta t$ in which Alice and Bob
switch the quadrature measurement is 5 ms in our experiment. The
long interval of 5 ms for switching the measurement bases was
limited by the phase locking technology of the homodyne detection
systems we held in the proof-of-principle experiment. Indeed, to
ensure the security the time interval $\Delta t$ should be as short
as possible which should be only confined by the storage time of
photons in the NOPA (It equals to the reciprocal of optical cavity
bandwidth. For our NOPA the minimal $\Delta t_{min}$ may reach $\sim
$ $5\times 10^{\text{-8}}$ s in principle.). Of course, we also can
enhance the security by lengthening the communication period to be
much longer than $\Delta t$ \cite{Lam}. We choose the sideband
frequency of $\Omega $ = 2 MHz as the centre frequency for Alice's
and Bob's measurements because the highest entanglement is obtained
at this frequency with our system. The measured initial correlation
variances of the amplitude
sum and phase difference between the signal and idler beams from the NOPA%
\textbf{\ }were 3.08 dB and 3.01 dB below the corresponding
shot-noise-level (SNL) at 2 MHz, which corresponding to the
correlation parameter $r=0.355$\ and $r=0.347$\ for amplitude and
phase quadratures, respectively. The output photocurrent from the
negative power combiner (-) passes through a low-noise amplifier of
30 dB and a band-pass filter (BPF) with the central frequency of 2
MHz and the bandwidth of 600 kHz, then it is recorded by a storage
oscilloscope (Agilent 54830B) at Alice and Bob, respectively. The
recorded data are transferred to a computer for the further data
processing. Before the communication, Alice and Bob should
synchronize their clocks and agree
on the time interval $\Delta t$ and the instantaneous measurement time $%
\Delta T$ in which a data point is taken. Only when the
instantaneous measurement time $\Delta T$ is longer than the storage
time of the NOPA, the measured correlations between the quadratures
of signal and idler beams have a stable value \cite{Rey}. The sample
rate of 10 MHz was chosen in the experiment. Since a band-pass
filter is placed before the storage oscilloscope in order to extract
the highly correlated quantum fluctuations, the real communication
bandwidth is reduced. Thus, in the data processing we digitally
re-sample the recorded data at 2 MHz (which corresponds to $\Delta
T=5\times 10^{\text{-7}}$ s).

\begin{table*}[tbp]
\caption{Experimental results for the different stages of the QKD
protocol used to distill the final secret key. Each step shows Alice
and Bob's mutual information ($I_{AB}$ bits/symbol), Eve's
information ($\protect\chi $ bits/symbol for collective attack and
$I_{AE}$ bits/symbol for individual attack), the corresponding net
information rate ($K$ bits/symbol and $\Delta I$ bits/symbol) and
the secret key rate (kbits/second) for 80\% and 40\%
transmission efficiency, respectively.}%
\par
\begin{center}
\begin{tabular}{|l|c|c|c|c|c|c|c|c|}
\hline & \multicolumn{8}{|c|}{80\% Transmission Efficiency} \\
\cline{2-9} & \multicolumn{4}{|c|}{Collective Attack} &
\multicolumn{4}{|c|}{Individual Attack} \\ \cline{2-9} & $I_{AB}$ &
$\chi $ & $K$ & Rate & $I_{AB}$ & $I_{AE}$ & $\Delta I$ & Rate
\\ \hline
Raw Data & 0.36 & 0.35 & 0.01 & 2000 & 0.38 & 0.23 & 0.15 & 2000 \\
\hline
Post-selection & 0.64 & 0.44 & 0.20 & 508 & 0.52 & 0.26 & 0.26 & 679 \\
\hline Reconciliation & $\sim 1$ & 0.68 & 0.32 & 346 & $\sim 1$ &
0.51 & 0.49 & 436
\\ \hline
Privacy Amplification & $\sim 1$ & $\sim 0$ & $\sim 1$ & 84 & $\sim 1$ & $%
\sim 0$ & $\sim 1$ & 109 \\ \hline & \multicolumn{8}{|c|}{40\%
Transmission Efficiency} \\ \cline{2-9} &
\multicolumn{4}{|c|}{Collective Attack} &
\multicolumn{4}{|c|}{Individual Attack} \\ \cline{2-9} & $I_{AB}$ &
$\chi $ & $K$ & Rate & $I_{AB}$ & $I_{AE}$ & $\Delta I$ & Rate
\\ \hline
Raw Data & 0.18 & 0.44 & -0.26 & 2000 & 0.18 & 0.33 & -0.15 & 2000
\\ \hline Post-selection & 0.69 & 0.63 & 0.06 & 46 & 0.44 & 0.35 &
0.09 & 180 \\ \hline Reconciliation & $\sim 1$ & 0.89 & 0.11 & 34 &
$\sim 1$ & 0.80 & 0.20 & 115
\\ \hline
Privacy Amplification & $\sim 1$ & $\sim 0$ & $\sim 1$ & 3 & $\sim 1$ & $%
\sim 0$ & $\sim 1$ & 10 \\ \hline
\end{tabular}%
\end{center}
\par
\end{table*}

After having recorded a string of data which include many data sets
(each set is measured within a phase switching time interval of 5
ms), Alice and Bob communicate through an authenticated public
channel to discard the sets measured on the incompatible bases and
to remain the compatible measured sets corresponding to the same
bases. The measured normalized variances of Alice's and Bob's
amplitude quadratures are 6.78$N_{0}$ and 7.02$N_{0}$ for 80\%
transmission efficiency, respectively. Considering the influence of
transmission efficiency, the variance measured by Alice is
$V_{A^{^{\prime
}}}=(\eta V_{A}+1-\eta )N_{0}=6.78N_{0}$ for the transmission efficiency $%
80\%$, so we have $V_{A}=8.23N_{0}$. Since $V_{A}=(V-1/V)N_{0}$ and $%
V_{s}=N_{0}/V$, we obtained $V=8.35N_{0}$ and $V_{s}=0.12N_{0}$. From $%
V_{B}=(\eta V_{A}+\eta V_{s}+1-\eta +\delta )N_{0}$ and Bob's
variance value we obtained the corresponding excess noise of $\delta
_{1}=0.14N_{0}$ for 80\% transmission efficiency. In the same way,
from the normalized variances of Alice's and Bob's phase quadratures
3.89$N_{0}$ and 4.05$N_{0}$ for 40\% transmission efficiency, we
calculated the excess noise of $\delta _{2}=0.11N_{0}$. Fig. 2 shows
the \textquotedblleft global\textquotedblright\ perspective of
Alice's and Bob's results measured on compatible bases, fig. 2 (a)
shows the function of the amplitude quadratures ($X_{B}$ vs $X_{A}$)
for 80\% transmission efficiency and fig. 2 (b) shows that of the
phase quadratures ($Y_{B}$ vs $Y_{A}$) corresponding to 40\%
transmission efficiency. The anti-correlation of the amplitude
quadratures ($\pm X_{a}\sim \mp X_{b}$) and the correlation of the
phase quadratures ($\pm Y_{a}\sim \pm Y_{b}$) are clearly exhibited
in the perspective. The quadrature measurements are normalized to
the SNL of the measured beam. Each one of fig. 2 (a) and (b)
contains 50,000 data points.

For extracting the secure data of $I_{AB}>\chi $ ( $I_{AB}>I_{AE}$)
from the measured raw data in the CV QKD, we used a post-selection
technique, that is, to select a subset from the measured raw data
points to make the mutual information of Alice and Bob advantage
over Eve's information. According to the way described before, Alice
and Bob select the secure data and discard the insecure data. The
dashed hyperbolas I and II in fig. 2 correspond to the secure
boundaries for collective attack and individual attack,
respectively. The regions at the outside of the hyperbolas I (II)
are secure $K$ $>$ 0 ($\Delta I$ $>$ 0) for the collective
(individual) attack, while the regions between the hyperbolas are
insecure (red points), the data in which should be discarded. The
green data points correspond to error-free bits, whilst the blue
data points correspond to that with bit-flip errors.

After post-selection procedure, Alice and Bob interpret the
post-selected data into binary data according to the way described
above. Then we apply the \textquotedblleft
Cascade\textquotedblright\ reconciliation protocol \cite{Bra} to
correct the errors of the retained data. At the stage of the error
correction, the data are arranged into many random subsets and the
error data are corrected. The efficiency of reconciliation is about
80\%. At last, we apply a privacy amplification procedure based on
universal hashing functions to distill the final secret key
\cite{Bennett,Cac}. First, Alice and Bob calculate a conservative
upper bound for Eve's knowledge about their key, then Alice and Bob
compute the parities of random subsets of the error-corrected key
bits. The obtained parity bits are kept as the final secret key. The
results for different stages of the QKD protocol used to distill the
secret key are shown in Table 1. The cost of these secret key
distillation processes is a reduction in the size of the secret key.
With the existence of the Gaussian collective attack, after the
privacy amplification procedure the final secret key rates of 84
kbits/s and 3 kbits/s are obtained for the transmission efficiencies
of 80\% and 40\%, respectively. To the Gaussian individual attack
only, the final secret key rates of 109 kbits/s and 10 kbits/s are
obtained for the transmission efficiencies of 80\% and 40\%,
respectively.

In conclusion, we accomplished the first experimental demonstration
of CV QKD protocol using the bright EPR entangled optical beams. The
quantum entanglement between two beams and the random quantum
fluctuations of amplitude and phase quadratures of respective
optical mode provide the physical mechanism for the CV QKD protocol
without the signal modulation. The security of the EB CV QKD
protocol against Gaussian collective and individual attack using
post-selection technique is analyzed. Although, as an example, the
binary coding scheme is utilized for simplification, Alice and Bob
can agree on a higher dimensional coding by dividing their results
into intervals corresponding to more than two bits values, in
principle. The presented CV QKD experiment intuitionally and
directly demonstrated the importance of the quantum entanglement for
the secure communication. It is possible to develop the more
complicated CV QKD networks by using the multipartite CV optical
entangled states based on this demonstrated scheme.

\acknowledgments This research was supported by the NSFC (Grants No.
60736040, 10674088, 10804065 and 60608012), NSFC Project for
Excellent Research Team (Grant No. 60821004), National Basic
Research Program of China (Grant No. 2006CB921101) and Shanxi
Province Science Foundation for Youths (Grant No. 2008021002).


\begin{thebibliography}{99}
\bibitem{Gisin} \Name{Gisin N., Ribordy G., Tittel W. \and Zbinden H.} %
\REVIEW{Rev. Mod. Phys.}{74}{2002}{145}.

\bibitem{Gros1} \Name{Grosshans F. et al.} \REVIEW{Nature}{421}{2003}{238}.

\bibitem{Lorenz} \Name{Lorenz S., Korolkova N. \and Leuchs G.}
\REVIEW{Appl. Phys. B}{79}{2004}{273}.

\bibitem{Lance} \Name{Lance A. M. et al.}
\REVIEW{Phys. Rev. Lett.}{95}{2005}{180503}.

\bibitem{Lod} \Name{Lodewyck J. et al.}
\REVIEW{Phys. Rev. A}{76}{2007}{042305}.

\bibitem{Qi} \Name{Qi B., Huang L. L., Qian L. \and Lo H. K.}
\REVIEW{Phys. Rev. A}{76}{2007}{052323}.

\bibitem{Got} \Name{Gottesman D., \and Preskill J.}
\REVIEW{Phys. Rev. A}{63}{2001}{022309}.

\bibitem{Ibl} \Name{Iblisdir S., Van Assche G., \and Cerf N. J.} %
\REVIEW{Phys. Rev. Lett.}{93}{2004}{170502}.

\bibitem{Gros2} \Name{Grosshans F.}
\REVIEW{Phys. Rev. Lett.}{94}{2005}{020504}.

\bibitem{Nav1} \Name{Navascu\'{e}s M., \and Ac\'{i}n A.}
\REVIEW{Phys. Rev. Lett.}{94}{2005}{020505}.

\bibitem{Ren} \Name{Renner R., \and Cirac J. I.}
\REVIEW{Phys. Rev. Lett.} {102}{2009}{110504}.

\bibitem{Lev} \Name{Leverrier A. \and Grangier P.}
\REVIEW{Phys. Rev. Lett.} {102}{2009}{180504}.

\bibitem{Curty} \Name{Curty M., Lewenstein M. \and L\"{u}tkenhaus N.} %
\REVIEW{Phys. Rev. Lett.}{92}{2004}{217903}.

\bibitem{Ralph} \Name{Ralph T. C.} \REVIEW{Phys. Rev. A}{61}{1999}{010303(R)}%
.

\bibitem{Hil} \Name{Hillery M.} \REVIEW{Phys. Rev. A}{61}{2000}{022309}.

\bibitem{Cerf} \Name{Cerf N. J., Levy M., \and Van Assche G.}
\REVIEW{Phys. Rev. A}{63}{2000}{052311}.

\bibitem{Reid} \Name{Reid M. D.} \REVIEW{Phys. Rev. A}{62}{2000}{062308}.

\bibitem{Ben} \Name{Bencheikh K., Symul TH., Jankovic A. \and Levenson J. A.}
\REVIEW{J. Mod. Opt.}{48}{2001}{1903}.

\bibitem{Sil} \Name{Silberhorn Ch., Korolkova N. \and Leuchs G.} %
\REVIEW{Phys. Rev. Lett.}{88}{2002}{167902}.

\bibitem{Su} \Name{Su X. L., Jing J. T., Pan Q. \and Xie C. D.} %
\REVIEW{Phys. Rev. A}{74}{2006}{062305}.

\bibitem{Pir} \Name{Pirandola S., Mancini S., Lloyd S. \and Braunstein S. L.}
\REVIEW{Nature Physics}{4}{2008}{726}.

\bibitem{Jia}
\Name{Jia X. J., Su X. L., Pan Q., Gao J. R., Xie C. D. \and Peng K.
C.} \REVIEW{Phys. Rev. Lett.}{93}{2004}{250503}.

\bibitem{Li}
\Name{Li X. Y., Pan Q., Jing J. T., Zhang J., Xie C. D. \and Peng K.
C.} \REVIEW{Phys. Rev. Lett.}{88}{2002}{047904}.

\bibitem{Takei} \Name{Takei N. et al.}
\REVIEW{Phys. Rev. A(R)}{74}{2006}{060101}.

\bibitem{Gros4} \Name{Grosshans F. \and Grangier P.}
\REVIEW{Phys. Rev. Lett.}{88}{2002}{057902}.

\bibitem{Silber}
\Name{Silberhorn Ch., Ralph T. C., L\"{u}tkenhaus N. \and Leuchs G.}
\REVIEW{Phys. Rev. Lett.}{89}{2002}{167901}.

\bibitem{Namiki} \Name{Namiki R. \and Hirano T.}
\REVIEW{Phys. Rev. Lett.}{92}{2004}{117901}.

\bibitem{Gros3} \Name{Grosshans F., et al.}
\REVIEW{Quantum. Inf. Comput.}{3}{2003}{535}.

\bibitem{Heid} \Name{Heid M. \and L\"{u}tkenhaus N.}
\REVIEW{Phys. Rev. A}{73}{2006}{052316}.

\bibitem{Van} \Name{Van Loock P.} \REVIEW{Fortschr. Phys.}{50}{2002}{1177}.

\bibitem{Lam} \Name{Lamoureux L. P. et al.}
\REVIEW{Phys. Rev. A}{73}{2006}{032304}.

\bibitem{Rey} \Name{Reynaud S., Fabre C. \and Giacobino E.}
\REVIEW{J. Opt. Soc. Am. B}{4}{1987}{1520}.

\bibitem{Bra} \Name{Brassard G. \and Salvail L.}
\Book{Advances in
Cryptology-Eurocrypt'93} \Editor{T. Helleseth} \Vol{765} %
\Publ{Springer-Verlag, Berlin} \Year{1994} \Page{410-423}.

\bibitem{Bennett}
\Name{Bennett C. H., Brassard G., Cr\'{e}peau C. \and Maurer U. M.}
\REVIEW{IEEE Trans. Inf. Theory}{41}{1995}{1915}.

\bibitem{Cac} \Name{Cachin C. \and Maurer U. M.}
\REVIEW{J. Cryptology}{10}{1997}{97}.
\end{thebibliography}
\end{document}